\documentclass[twocolumn,showpacs,preprintnumbers,amsmath,amssymb]{revtex4}

\begin{document}
\draft
\title{On the ground state and excitations of a four-component fermion model}
\author{You-Quan Li$^{1,2}$, Guan-Shan Tian$^{1,3}$, Michael Ma$^{1,4}$,
        and Hai-Qing Lin$^{1}$}
\address{
$^{1}$Department of physics, Chinese University of Hong Kong, Hong Kong, China\\
$^{2}$Zhejiang Institute of Modern Physics, Zhejiang University,
    Hangzhou 310027, China\\
$^{3}$Department of Physics, Peking University, Beijing, China\\
$^{4}$Department of Physics, University of Cincinnati, Cincinnati,
Ohio, 45221,
USA\\
}
\date{Received: \today }

\begin{abstract}
The properties of the ground state and excitations of a
four-component Hubbard-like model are studied by the
Lieb-Schultz-Mattis approach. It is proven rigorously that the
ground state of the model is nondegenerate and excitations are
gapless at all band fillings.
\end{abstract}



\maketitle


The nature of the ground state and excitations of strongly
correlated electronic systems plays an important role in our
understanding of various fascinating phenomena such as
metal-insulator transition, high-temperature superconductivity and
colossal magnetoresistance. Recently, there has been much
interests in the studies of correlated electrons with orbital
degree of freedom\cite{Nagaosa}. In the case of double orbital
degeneracy, a SU(4) theory was proposed as an ideal limit.
\cite{LiMSZ}.

Although there exists few rigorous results for general many-body
Hamiltonians, significant progress has been made in one spatial
dimension. In 1961, Lieb, Schultz, and Mattis (LSM) proved
\cite{LSM} a remarkable theorem: For the spin 1/2 Heisenberg
antiferromagnetic chain, the ground state is nondegenerate and the
energy spectrum is gapless. This theorem has been extended
recently to a hierarchy of generalized Heisenberg models \cite
{Afflieb,Li01} with high ranking symmetries, including the SU(4)
model. It is of interest to obtain results for systems where the
kinetic terms caused by nearest neighbor hopping is not
negligible. For the two-component Hubbard model, the
non-degeneracy of the ground state was shown by Lieb\cite{Lieb89}
and gaplessness at 1/2 filling by Yamanaka, Oshikawa, and Affleck
(YOA)\cite{OYAffleck97b}. The extension of these results to four
component Hubbard-type models is valuable because such models not
only describe electrons with double orbital degeneracy but also
two layer systems without inter-layer hopping. Experimental
realization of four component or two-band Hubbard model in one
dimension includes quasi one
dimensional materials such as Na$_{2}$Ti$_{2}$Sb$_{2}$O amd NaV$_{2}$O$_{5}$%
\cite{PSK}.

In this work, we consider a four-component Hubbard model defined
on a one-dimensional lattice with length $L$
\begin{eqnarray}
H=-t\sum_{\stackrel{a}{<x,x^{\prime
}>}}c_{a}^{+}(x)c_{a}(x^{\prime
})+U\sum_{\stackrel{a<a^{\prime }}{x}}n_{a}(x)n_{a^{\prime }}(x) \nonumber\\
+V\sum_{%
\stackrel{a,a^{\prime }}{<x,x^{\prime }>}}n_{a}(x)n_{a^{\prime
}}(x^{\prime })~,  \label{eq:Hamiltonian}
\end{eqnarray}
where $a,a^{\prime }=1,2,3,4$; $x$ identifies the lattice site,
and $\langle x,x^{\prime }\rangle $ stands for nearest neighbor
sites. $c_{a}^{+}(x)$
creates a fermion of component $a$ on site $x,$ and $%
n_{a}(x)=c_{a}^{+}(x)c_{a}(x)$ is the corresponding number
operator. The on-site interaction is assumed to be repulsive. The
four components can be related to electrons with doubly orbital
degeneracy. Let us define the spin components by up $(\uparrow )$
and down $(\downarrow )$ and the orbital components by top and
bottom. Then the four possible electron singly occupied states are
$|1\rangle =|
\begin{array}{c}
\uparrow  \\[-1mm]
-
\end{array}
\rangle ,\;|2\rangle =|
\begin{array}{c}
\downarrow  \\[-1mm]
-
\end{array}
\rangle \;|3\rangle =|
\begin{array}{c}
- \\[-1mm]
\uparrow
\end{array}
\rangle ,\;|4\rangle =|
\begin{array}{c}
- \\[-1mm]
\downarrow
\end{array}
\rangle $.

Previous literature on such systems has largely concentrated on
the strong coupling limit of this Hamiltonian, whereby it is
mapped onto a $SU(4)$
Heisenberg model\cite{Sutherland,YSU1} and related models away from $%
SU(4)$ symmetry\cite{YSU2,IQA}. For the $SU(4)$ Heisenberg
Hamiltonian, it has been established that there are gapless
excitations at crystal momentum $K=\pm \pi /2$ and $\pi
$\cite{LiMSZ,Sutherland,YSU1}$.$ In the case of no orbital
degeneracy, it is well known that due to the existence of the
charge gap for any $U>0,$ the low energy physics of the Hubbard
model is essentially identical to the $SU(2)$ Heisenberg model. It
is of interest to investigate if this is the case for the $SU(4)$
case also. In particular, much insight can be obtained if exact
and rigorous results are available to complement approximate
results using methods like bosonization \cite{Boson}.

In the following, we first prove that the ground state of the Hamiltonian (%
\ref{eq:Hamiltonian}) with periodic boundary condition is
nondegenerate when the total number of electrons is $N=4n$ with
$n$ being an odd integer. We then study whether the excitations
are gapless for different band filling.

To show nondegeneracy of the ground state, we shall apply a
simplified version of Lieb and Mattis' method in proving the
absence of ferromagnetism in one-dimensional itinerant electron
lattice models \cite {Lieb-Mattis,Tian-Lin}. First, we notice
that, according to the representation theory of Lie algebra
$A_{3}$, for any multiplet, there is always one state lying in the
subspace of zero weight $(0,\>0,\>0)$, which is spanned by the
vectors with equal number of different kinds of fermions.
Therefore, by introducing operators $\hat{{\cal C}}_{a}^{\dagger }({\bf x}%
)=c_{a}^{\dagger }(x_{1})c_{a}^{\dagger }(x_{2})\cdots
c_{a}^{\dagger }(x_{n})$ with $x_{1}<x_{2}<\cdots <x_{n}$, we can
write the ground state wavefunction as
\begin{equation}
|\psi _{0}\rangle =\sum_{\mu }W_{\mu }\hat{{\cal C}}_{1}^{\dagger }({\bf x})%
\hat{{\cal C}}_{2}^{\dagger }({\bf x}^{\prime })\hat{{\cal
C}}_{3}^{\dagger }({\bf x}^{\prime \prime })\hat{{\cal
C}}_{4}^{\dagger }({\bf x}^{\prime \prime \prime })\left|
0\right\rangle ,  \label{Wave Function}
\end{equation}
where $\mu $ stands for the set $\{{\bf x},\>{\bf x}^{\prime },\>{\bf x}%
^{\prime \prime },\>{\bf x}^{\prime \prime \prime }\}$. Obviously,
$\psi _{0} $ belongs to the subspace $V_{0}(n)$, in which each
vector has the same number $n$ of $a=1,\>2,\>3$ and $4$ types of
fermions. The total number of expansion terms in Eq.~(\ref{Wave
Function}) is $\left( C_{L}^{n}\right) ^{4}$.

In terms of this basis of vectors, we are able to write
Hamiltonian (\ref {eq:Hamiltonian}) into a matrix ${\cal H}$ which
has the following
characteristics: (i) All of its off-diagonal elements are either zero or $-t$%
, although its diagonal elements may have different signs. The
reason why off-diagonal elements all have the same sign is due to
the fact that, in Hamiltonian (\ref{eq:Hamiltonian}), only hopping
terms between nearest-neighbor sites are present. In one
dimension, such hopping will not change the positional order of
the fermions except at the boundary. However, if periodic boundary
condition is used for $n$ odd, then the boundary hopping will not
cause sign problems. (ii) Furthermore, ${\cal H}$ is also
irreducible. Namely, for any pair of indices $m$ and $n$, there is
a
positive integer $M$ such that the matrix element $\left( {\cal H}%
^{M}\right) _{mn}$ is nonzero since the chain is connected by
fermion hopping. Therefore, any pair of configurations $\hat{{\cal
C}}_{1}^{\dagger
}({\bf x})\hat{{\cal C}}_{2}^{\dagger }({\bf x}^{\prime })\hat{{\cal C}}%
_{3}^{\dagger }({\bf x}^{\prime \prime })\hat{{\cal C}}_{4}^{\dagger }({\bf x%
}^{\prime \prime \prime })|0\rangle $ and $\hat{{\cal C}}_{1}^{\dagger }(%
{\bf y})\hat{{\cal C}}_{2}^{\dagger }({\bf y}^{\prime })\hat{{\cal C}}%
_{3}^{\dagger }({\bf y}^{\prime \prime })\hat{{\cal C}}_{4}^{\dagger }({\bf y%
}^{\prime \prime \prime })|0\rangle $ in subspace $V_{0}(n)$ are
connected by an appropriate number $M$ of fermion hoppings.

To such a Hermitian matrix, we are able to apply the well-known
Perron-Fr\"obenius theorem \cite{Franklin}. It tells us that {\it
the expansion coefficients in Eq.~(\ref{Wave Function}) have the
same sign and hence, $\psi_0$ is nondegenerate in subspace
$V_0(n)$}. To show that it is globally nondegenerate, we need to
determine its quantum numbers by the continuity argument.
Adiabetically turning off the interactions of Hamiltonian
(\ref{eq:Hamiltonian}), we reduce it to a tight-binding $H_0$,
whose ground state in $V_0(n)$ is a spin-orbital singlet. It implies that $%
\psi_0$ is also a spin-orbital singlet and hence, is globally
nondegenerate.

Next we study the nature of excitations for the case of $N=4n$
with $n$ odd, focusing on whether they are gapless. Following LSM
\cite{LSM} and YOA\cite {OYAffleck97b}, we consider the state
\begin{equation}
|\psi _{P}\rangle =\exp \biggl[ -i\frac{2\pi }{L}\sum_{x=1}^{L}xP(x)\biggr] %
|\psi _{0}\rangle \equiv \hat{O}(P)|\psi _{0}\rangle ~,
\label{eq:twist}
\end{equation}
where operator $P(x)$ $=\sum_{\alpha }$ $v_{\alpha }n_{\alpha
}(x),$ and the coefficients $v_{\alpha }$ are to be chosen by the
reasoning below. Our aim is to prove gaplessness. In order to do
so, we want $|\psi _{P}\rangle $ to be orthogonal to $|\psi
_{0}\rangle $ and to have an energy expectation value which equals
the ground state energy in the limit $L\rightarrow \infty.$ The
operator $\hat{O}(P)$ ``boosts'' the crystal momentum of each
$\alpha $ component particle by $2 \pi v_{\alpha }/L.$ In order to
preserve periodic boundary condition, $v_{\alpha }$ has to be an
integer. The state $|\psi _{P}\rangle $ then has a crystal
momentum of $K=\frac{2\pi}{L}\sum_{x=1}^{L}P(x)$ relative to that
of the ground state. Since the Hamiltonian is translationally
invariant, and the ground state is non-degenerate, it has a
definite crystal momentum. Thus, a sufficient condition for the
orthogonality condition is $K \neq 2m \pi,$ where $m$ is any
integer, i.e., $K$ should not be a reciprocal lattice vector. The
expectation energy of $|\psi _{P}\rangle $ is given by
\[
E_{P}=\langle \psi _{P}|H|\psi _{P}\rangle =\langle \psi
_{0}|H^{\prime }|\psi _{0}\rangle
\]
where
\[
H^{\prime }=\hat{O}^{\dagger }H\hat{O} ~.
\]
Since the interacting part of $H$ commutes with $\hat{O}$ the only
difference between $H^{\prime }$ and $H$ is in the tight binding
part, and
is given by a  phase rotation of each $c_{\alpha }(x)\rightarrow e^{-i2\pi %
v_{\alpha }x/L}c_{\alpha }(x).$  This results in a change of the
tight binding amplitude
\[
t\rightarrow t_{x,x+1}^{(\alpha )}=te^{-i2\pi v_{\alpha }/L},\quad
x\neq L
\]
and $t_{L,1}^{(\alpha )}=te^{i2\pi v_{\alpha }(L-1)/L}.$ Because
$v_{\alpha }$ is chosen to be an integer, periodic boundary
condition is preserved, and this last bond will remain equivalent
to all the other bonds. As a result, the difference $H^{\prime
}-H$ will vanish as $L\rightarrow \infty .$ More precisely, using
$P(x)=n_{1}(x)$ as an example, we have
\begin{eqnarray*}
E_{P}-E_{0} &=&-t(e^{-i2\pi /L}-1)\sum_{x}\langle c_{1}^{\dagger
}(x)c_{1}(x+1)\rangle +c.c. \\
&=&-2t(\cos \frac{2\pi }{L}-1)\sum_{x}\langle c_{1}^{\dagger
}(x)c_{1}(x+1)\rangle
\end{eqnarray*}
where the last line follows from $\langle c_{1}^{\dagger
}(x)c_{1}(x+1)\rangle $ being real for a non-degenerate ground
state. Actually, this method can only tell us that there is at
least one low lying state with energy at most of order $1/L$
higher than the ground state. Thus, the ``gaplessness'' may be due
to having true gapless excitations or a discrete number of
symmetry breaking states which are degenerate in the thermodynamic
limit \cite{Afflieb,MTL}. For this reason, our results below
should all be taken implicitly as assuming there is no discrete
symmetry breaking such as lattice translation.

In particular we consider states generated by the following four
operators
\begin{eqnarray}
P_{1}(x) &=&n_{1}(x)~,  \nonumber \\
P_{2}(x) &=&n_{1}(x)+n_{2}(x)~,  \nonumber \\
P_{3}(x) &=&n_{1}(x)+n_{2}(x)+n_{3}(x)~,  \nonumber \\
P_{4}(x) &=&n_{1}(x)+n_{2}(x)+n_{3}(x)+n_{4}(x)~.
\label{4-operator}
\end{eqnarray}
For general band filling, the crystal momentum $K$ for all $4$
states will not be equal to a reciprocal lattice vector, and they
are all orthogonal to
the ground state, and there are gapless excitations at $K_{i}$ generated by $%
P_{i}$. Band fillings equal to $1/4~(N=L)$, $1/2~(N=2L)$, and
$3/4~(N=3L)$, however, must be discussed separately.

(i) 1/4-filled case: the ground state is a SU(4) singlet and we have $%
N_{1}=N_{2}=N_{3}=N_{4}=N/4=L/4$, where $N_{a}=\sum_{x}n_{a}(x)$.
The
crystal momenta of the $4$ states given by Eq. (\ref{4-operator}) are then $%
\pi /2,$ $\pi,$ $3\pi /2,$ and $2\pi $ respectively. Thus all $4$
states are mutually orthogonal to each other, with the first three
also orthogonal to the ground state. The fourth state cannot be
concluded to be orthogonal to the ground state. Note that except
for the non-interacting limit, the crystal momentum for each
flavor is not a conserved quantity, only the total crystal
momentum. Therefore, other combinations like $2n_{1}$ for example
cannot be shown to be orthogonal to $n_{1}+n_{2}.$

Our analysis establishes that at least at momenta $\pi /2,$ $\pi
,$ $3\pi /2\equiv -\pi /2,$ there exists gapless excitations. In
the large $U$ limit, where the 4-component Hubbard model is mapped
into an $SU(4)$ symmetric
Heisenberg model, it is known exactly that there are gapless excitations at $%
K=0,\pi /2,$ $\pi,$ and $3\pi /2$ and nowhere else. While our
modified LMS approach cannot say anything about $K=0$ and cannot
rule out gapless excitations at other $K,$ it supports that the
gapless modes are pinned at these $K$ values for all $U.$

(ii) 3/4-filled case: this case is equivalent to the 1/4-filled
case by particle-hole symmetry.

(iii) 1/2-filled case: the ground state is again a singlet and we have $%
N_{1}=N_{2}=N_{3}=N_{4}=N/4=L/2$. The 4 states from Eq.
(\ref{4-operator}) now have momenta $K=\pi, 2\pi, 3\pi $, and
$4\pi.$ Thus we can establish that gapless excitations must exist
at $K=\pi.$

While gapless excitations are conclusively shown by the above
analysis, they do not tell us much about the excitations beyond
the crystal momentum. To try to understand these excitations more,
we first consider the single band Hubbard model. By similar
approach, one can show that twisted wavefunctions, for both spin
up and down,
\begin{equation}
|\psi _{n_{\sigma }}\rangle =\exp \biggl[ -i\frac{2\pi }{L}%
\sum_{x=1}^{L}xn_{\sigma }(x)\biggr] |\psi _{0}\rangle ~,
\end{equation}
have crystal momentum $K=\pi $ at 1/2 filling and are orthogonal
to the ground state. In the non-interacting limit, crystal
momentum for each spin is conserved, and $\langle \psi
_{n_{\uparrow }}|\psi _{n_{\downarrow }}\rangle =0,$ so there are
two distinct gapless modes. In this limit, the
gapless excitation with $K=\pi $ is a single particle-hole excitation of $%
2k_{F}$ for one of the spins$.$ These can then be cast in spin
symmetric combination and spin antisymmetric combinations to give
gapless charge and spin excitations. For non-zero $U,$ the Hubbard
model at 1/2 filling always has a charge gap and only spin
excitations are gapless. Correspondingly, only total crystal
momentum is conserved and $|\psi _{n_{\uparrow }}\rangle $ and
$|\psi _{n_{\downarrow }}\rangle $ no longer have to be
orthogonal. Indeed, their overlap, $\langle \psi _{n_{\uparrow
}}|\psi _{n_{\downarrow }}\rangle $ is nonzero as verified by
numerical calculations. Rewrite
\[
n_{\uparrow }=\frac{1}{2}(n_{\uparrow }+n_{\downarrow })+\frac{1}{2}%
(n_{\uparrow }-n_{\downarrow })=\frac{1}{2}n+S_{z}~,
\]
one has
\begin{eqnarray*}
\exp \biggl[ -i\frac{2\pi
}{L}\sum_{x=1}^{L}xn_{\uparrow
}(x)\biggr] = \exp \biggl[
-i\frac{\pi
}{L}\sum_{x=1}^{L}xn(x)\biggr]\times
\\\exp \biggl[ -i\frac{2\pi
}{L}\sum_{x=1}^{L}xS_{z}(x)\biggr]
=\hat{O}(n)\hat{O}(S_{z})~.
\end{eqnarray*}
{\it \ } In the strong
interaction limit,
$U\rightarrow \infty $,each
site tends to be singly
occupied, $n(x)\rightarrow
1$, so what left is the
spin
fluctuation $S_{z}(x)$. 
For finite $U$, it would seem that $n_{1}$ generates both charge
and spin excitations. However, in order for gaplessness, it must
be that the overlap with charge excitations must vanish {\it in
the thermodynamics limit}. While we cannot show this rigorously,
we can get a partial understanding by considering the single mode
approximation state
\[
|\psi _{1}\rangle =\sum_{x=1}^{L}\exp \left( i\pi x\right)
n_{1}(x)|\psi
_{0}\rangle =\frac{1}{2}\widetilde{n}(\pi )|\psi _{0}\rangle +\widetilde{S}%
_{z}(\pi )|\psi _{0}\rangle ~.
\]
So $\langle \psi _{1}|\psi _{1}\rangle =\frac{1}{4}\left\langle \widetilde{n}%
(-\pi )\widetilde{n}(\pi )\right\rangle +\left\langle
\widetilde{S}_{z}(-\pi )\widetilde{S}_{z}(\pi )\right\rangle .$
Due to the charge gap, the density-density correlation is finite
while the spin-spin correlation
diverges as $L\rightarrow \infty .$ Thus there is no spectral weight for $%
|\psi _{1}\rangle $ to be in a charge excitation. Conversely, our
proof of gaplessness at $\pi $ shows that the spin-spin
correlation function must diverge at $\pi.$

The twisted operator $\exp \biggl[ -i\frac{2\pi }{L}\sum_{x=1}^{L}xS_{z}(x)%
\biggr]$, is what Lieb, Schultz, and Mattis used in their paper
(Ref. \cite {LSM}) to show gapless excitation in the Heisenberg
model. It cannot be shown to be orthogonal to the ground state in
the Hubbard model because it
has crystal momentum $0$. Instead $n_{\sigma }(x)$ must be used instead of $%
S_{z}(x).$ The fact that $\hat{O}(n_{\sigma })$ gives a stronger
result than that $\hat{O}(n_{\uparrow }+n_{\downarrow })$ and
$\hat{O}(n_{\uparrow }-n_{\downarrow })$ was noted by Yamanaka,
Oshikawa, and Affleck \cite {OYAffleck97b}.

Returning to the 4-component Hubbard model, we discuss the
analogous situation for the case of $N=L/4.$ In this case, our
proof shows gapless excitations for $\pm \pi /2$ and $\pi.$ In the
non-interacting limit, there
are 4 orthogonal choices for $P_{1}$ corresponding to whether we use $%
n_{1,}n_{2,}n_{3,}$ and $n_{4},$ and they generate a single
particle-hole excitation of $2k_{F}$. Linear combinations of these
can be taken to form charge, spin, orbital, and spin-orbital
excitations, the last three corresponding to the $3$ diagonal
generators of $SU(4).$ For $U\neq 0,$ the four $n_{\alpha }$
states are not orthogonal, and whether there are $3$ or $4 $
gapless modes will depend on whether there is a charge gap (the
other three are related by symmetry so they must be all gapless)
\cite{Boson}. The ``pinning'' of gapless excitations at $K=2k_{F}$
of the non-interacting system was used by YOA as a generalized
definition of Luttinger's Theorem in 1D. In analogy to the
2-component Hubbard model, the gaplessness implies the spin-spin
correlation function, the orbital-orbital correlation functions,
and the spin-orbital-spin-orbital correlation functions must diverge at $%
K=\pi /2.$ In the strong coupling limit, charge fluctuations are
frozen, and we are definitely left with $3$ gapless excitations,
which are the three
states with $N_{1}=N_{2}=N_{3}=N_{4}=N/4$ in the $15$ representation of \ $%
SU(4).$ For $K=\pi =4k_{F},$ the gapless excitations for the
non-interacting systems are pairs of $2k_{F}$ particle-hole
excitations. The correlation functions that diverge at $4k_{F}$
are those corresponding to 4-particle Green's functions.

For filling factor $N_{\alpha }=p/q,$ with $p$ and $q$ integers,
the
orthogonality condition that $K\neq 2m\pi $ will be satisfied for $%
\sum_{\alpha }v_{\alpha }=1,2,...,q-1.$ Accordingly there will be
at least $q$
momenta of gapless excitations. In the non-interacting limit, the $%
\sum_{\alpha }v_{\alpha }=r$ gapless excitations are $r$
particle-hole pairs. If we assume no broken translational
invariance due to interactions (which should be the case for the
Hubbard model with only on site repulsion), then gapless
excitations at these momenta persist when $U\neq 0$ \cite
{OYAffleck97b}.

In summary, we studied properties of the ground state and
excitations of the four-component Hubbard model, in which
electrons carry spin as well as orbital degrees of freedom.
Considering cases where total number of electrons $N=4n$ with $n$
odd for periodic boundary condition,
we showed rigorously that the ground state of this system is
nondegenerate. Using the twist operators as specified in Eq. (\ref
{4-operator}), we addressed the issue of the existence of gapless
excitations. We showed that away from the filling factor 1/4, 1/2,
and 3/4, the state produced by acting the twist operators (e.g.,
Eq. (\ref{4-operator})) on the nondegenerate ground state is
orthogonal to the ground state and its variational energy
approaches to the ground state energy in the thermodynamic limit.
For the filling factor equals to 1/4 and 3/4, we showed that
gapless excitations exist at crystal momenta $\pi /2,$ $\pi ,$
$3\pi /2\equiv -\pi /2$.

The work is supported by
the Research Grants Council
of Hong Kong under projects
4246/01P and C001 PHY. YQL
acknowledges support by
NSFC No. 10225419 and No.
90103022. GST acknowledges
support by NSFC.


\end{document}